\documentclass[10pt]{article}

\usepackage[OE]{express}
\usepackage[colorlinks=true,urlcolor=blue,linkcolor=black,citecolor=black,breaklinks=true,bookmarks=false]{hyperref}
\usepackage{graphicx}
\usepackage{siunitx} 
\usepackage{tabularx}
\usepackage{soul} 
\usepackage{amsmath}
\usepackage{lipsum}
\usepackage{cite}
\usepackage{lmodern}
\usepackage{tablefootnote}

\begin{document}
\title{Characterization of high-temperature performance of cesium
  vapor cells with anti-relaxation coating}

\author{Wenhao~Li,$^{1,2}$ Mikhail~Balabas,$^3$ Xiang~Peng,$^1$ Szymon~Pustelny,$^4$ Arne~Wickenbrock,$^5$ Hong Guo,$^{1,*}$ and Dmitry~Budker$^{2,6,7,**}$}

\address{
  $^1$State Key Laboratory of Advanced Optical Communication Systems and Networks, School of Electronics Engineering and Computer Science, and Center for Quantum Information Technology, Peking University, Beijing 100871, China\\
  $^2$Department of Physics, University of California, Berkeley, CA 94720-7300\\
  $^3$St. Petersburg State University, 7/9 Universitetskaya nab., St. Petersburg 199034, Russia\\
  $^4$Institute of Physics, Jagiellonian University, Lojasiewicza 11, 30-348 Krak\'{o}w, Poland\\
  $^5$Johannes Gutenberg-University Mainz, 55128 Mainz, Germany\\
  $^6$Helmholtz Institute Mainz, 55099 Mainz, Germany\\
  $^7$Nuclear Science Division, Lawrence Berkeley National Laboratory, Berkeley, CA 94720}

\email{$^*$hongguo@pku.edu.cn}
\email{$^{**}$budker@uni-mainz.de} 


\begin{abstract}
Vapor cells with antirelaxation coating are widely used in modern atomic physics experiments due to the coating's ability to maintain the atoms' spin polarization during wall collisions. We characterize the performance of vapor cells with different coating materials by measuring longitudinal spin relaxation and vapor density at temperatures up to $\SI{95}{\degreeCelsius}$. We found that the spin-projection-noise-limited sensitivity for atomic magnetometers with such cells improves with temperature, which demonstrates the potential of antirelaxation coated cells in applications of future high-sensitivity magnetometers.
\end{abstract}



\section{Introduction}

Alkali metal atoms with long-lived ground-state spin polarization play an important role in a wide range of applications, such as atomic magnetometry\cite{budker2007}, frequency standards\cite{Budker2005}, quantum memory for light\cite{Julsgaard2004}, generation of spin-squeezed states\cite{Kuzmich2000}, and sensitive searches for exotic physics\cite{Pustelny2013}. Two common methods to preserve the spin state are the use of chemically inert buffer gas and antirelaxation wall coating (AWC). Advantages of magnetometers based on coated cells over those based on buffer-gas cells include the ability to operate at lower light power, reduced sensitivity to magnetic-field gradients\cite{Pustelny2006}, and narrower spectral lines\cite{Bouchiat1966,Liberman1986,Graf2005,Karaulanov2009,Balabas2010,Balabas2010a,Balabas2012,Nasyrov2015,Seltzer2010}. High-quality AWC can allow for up to $10^6$ bounces of a polarized alkali atom off the cell wall without depolarization\cite{Balabas2010PRL}.

The high-temperature operation of AWC is important for applications requiring a high atomic vapor density, such as miniaturized magnetometers\cite{Jensen2016}, atomic clocks based on miniaturized cells\cite{Kitching2002}, and spin-exchange-relaxation-free (SERF) magnetometers with low buffer gas pressure \cite{Ledbetter2008}. When heating vapor cells, the saturated vapor density increases, which can improve the sensitivity of spin-related measurements. On the other hand, heating can also deteriorate AWC's performance or cause the coating material to melt (for a typical paraffin AWC, the melting point is around $\SI{60}{\degreeCelsius}$). High-temperature operation up to about $\SI{170}{\degreeCelsius}$ was demonstrated with octadecyltrichlorosilane (OTS) coatings\cite{Seltzer2009}, however, the authors report typical performance in terms of depolarization rates to be about several hundred wall collisions without depolarization. In addition, significant variation of AWC properties was observed with a nominally uniform preparation procedure\cite{Seltzer2009}, which may be problematic for mass production of AWC cells. Although the properties of AWC were studied extensively \cite{Balabas2012,Nasyrov2015,Balabas2013} and some previous works specially focused on the relaxation properties of AWC near the melting point\cite{Bouchiat1966,Seltzer2010,Rahman1987,Vanier1981}, the effect of heating AWC cells above the coating-material melting point on the magnetometric sensitivity remains insufficiently explored.

For atomic magnetometers, the spin-projection-noise-limited (atomic-shot-noise-limited) sensitivity $\delta B_{\mathrm{SNL}}$ of a polarized atomic system to magnetic field can be described by\cite{budker2013optical}
\begin{equation}\label{equ:snl}
  \delta B_{\mathrm{SNL}} \approx \frac{1}{\gamma} \sqrt{\frac{1}{n\tau_{\mathrm{rel}} T}},
\end{equation}
where $\gamma$ is the gyromagnetic ratio, $\tau_{\mathrm{rel}}$ is the spin relaxation time \cite{footnote1}, $n$ is the number of measured atoms, and $T$ is the measurement time. From Eq.~(\ref{equ:snl}), we can see that once the atomic system is determined, the spin-projection-noise-limited sensitivity will essentially depend on the product of the number of measured atoms and the spin relaxation time. Therefore in this article, we demonstrate the results of systematic measurements of vapor density and spin relaxation time at high temperatures. For the vapor-density measurement, the optical transmission signal of a weak laser beam is fitted to the atomic absorption function. For the spin-relaxation-time measurement, in contrast to the traditional Franzen's ``relaxation in the dark'' (RID) method\cite{Franzen1959}, where the transmission of circularly polarized light is monitored, we use a modified method where circularly polarized pump light orients the atomic ensemble and the optical rotation of linearly polarized probe light is observed\cite{Graf2005}. The main advantage of this modified RID over the original version is that the signal only involves two time constants, which simplifies the characterization of vapor cells\cite{Graf2005} (Another version of the RID technique that enables measurements of the transverse spin relaxation is discussed in
\cite{Chalupczak2013}.)

\section{Experimental setup and measurement procedure}

\begin{figure}[thbp]
  \centering\includegraphics[]{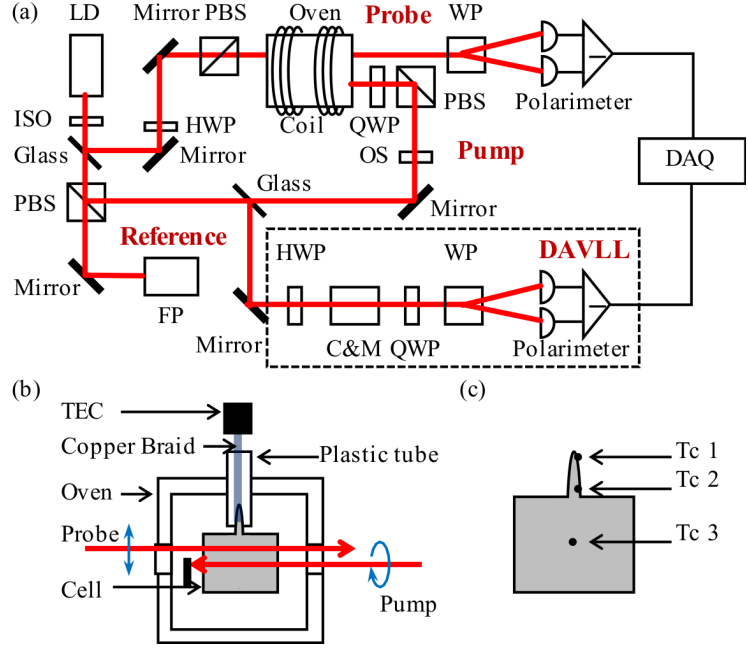}%
  \caption{\label{fig:setup} (a) Schematic of the experimental setup. LD: Laser diode; ISO: optical isolator; PBS: polarizing beam splitter; FP: Fabry-P\'erot spectrum analyzer; HWP: half-wave plate; QWP: quarter-wave plate; OS: optical shutter; C\&M: cell and magnet system; WP: Wollaston prism; DAQ: data acquisition system. (b) Schematic of the cell temperature control system. TEC: thermoelectric cooler. Twisted heating wires are not shown. (c) Schematic of the cell temperature measurement configuration. A cylindrical cell is shown as an example, spherical cells have a similar arrangement. The black dots show the three points where thermocouples (Tc) are attached to during the measurements. Tc~1 - Tc~3 measure the stem temperature, the capillary temperature, and the volume temperature, respectively. The volume temperature also represents the coating temperature.}%
\end{figure}

The experimental setup is shown in Fig.~\ref{fig:setup}. A pair of Helmholtz coils generates a magnetic field of around 6~G along the beams' propagation direction. The magnitude of the field should be larger than the ambient field in the lab (0.5~G), so that there is a well defined field axis and no unwanted spin precession, but also not too big to avoid effects due to polarization moments other than orientation\cite{Graf2005}. In our experiment, spin relaxation is essentially independent of the magnetic field in the range from 1~G to 10~G which validates the assumption. Probe beam and pump beam are generated by a 852~nm (Cs D2 line) distributed feedback (DFB) laser whose frequency is locked to the Doppler broadened $F_{\mathrm{g}}=4 \rightarrow F_{\mathrm{e}}=3,4,5$ transitions with a compact dichroic atomic vapor laser lock (DAVLL) system\cite{Yashchuk2000}. The DAVLL signal is used as the input of an analog feedback control loop (Stanford Research System SIM960). The broad DAVLL error signal allows stable laser frequency locks within a 2.6~GHz range centered around the $F_{\mathrm{g}}=4 \rightarrow F_{\mathrm{e}}=5$ transition \cite{footnote2}. The probe beam power is kept as low as $1~\mu W$ to avoid deformation of the optical spectrum due to saturated optical absorption. The pump beam power is about $15~\mu W$, which is sufficient to optically pump a significant number of atoms and therefore to generate Zeeman polarization in the $F=4$ hyperfine manifold. The diameters of the probe beam and the pump beam are about 1~mm and 5~mm, respectively, and are kept constant for all the measurements. And there is no cross-sectional area of the probe and the pump beam. To study spin relaxation transients the pump beam was switched on and off with a mechanical shutter (Thorlabs SH05). Pump beam and probe beam are counter-propagating to minimize pump-light scattering into the probe photodetectors.

We investigated several vapor cells, different in shape, volume, capillary size, and coating material. The capillary is the part that connects the cell stem and the cell and governs the exchange of atoms between the two volumes. The existence of the capillary connected to the main cell volume is the origin of reservoir relaxation\cite{Graf2005}. The parameters for the vapor cells used in this experiment are summarized in Table.~\ref{tab:cell}. pwMB is the same wax fraction investigated in \cite{Seltzer2010}. pwMB--H is the fraction of the same initial wax distilled at $\SI{240}{\degreeCelsius}$. Cell~E is coated with an alkene-based coating, which is similar to the coating used in \cite{Balabas2010PRL}. Cell~E has a glass ``lock'' in the stem to control atom diffusion between the volume and the stem. The ``lock'' is kept open during the measurements to ensure identical conditions with respect to the measurements with the other cells. Cell~F and cell~G are coated with commercial paraffin wax and are independently manufactured at the Peking University. Cell~H is coated with deuterated polyethylene, which has a higher melting temperature than other coating materials we investigated in the paper. It should be noted that the melting temperatures listed in Table.~\ref{tab:cell} are estimated values. More precise measurement of the melting point of the paraffin wax can be done with differential scanning calirometry (DSC) as described in \cite{Seltzer2010}.

\begin{table*}[thbp]
  \caption{Parameters for vapor cells used in the experiments. The fast relaxation rate $\gamma_{\mathrm{f}}$ and the slow relaxation rate $\gamma_{\mathrm{s}}$ shown in the table are measured at $T_{\mathrm{volume}} \approx \SI{30}{\degreeCelsius}$ and $T_{\mathrm{Stem}} \approx \SI{20}{\degreeCelsius}$. $d$ and $l$ is the diameter and the length of the cell, respectively. $N$ is the average number of wall collisions without depolarization calculated with slow relaxation rates. A detailed description of the spin-relaxation measurement is in Section.~\ref{sec:spin-relaxation}.}
  \centering
  \begin{tabular}{p{0.04\linewidth}p{0.12\linewidth}p{0.04\linewidth}p{0.04\linewidth}p{0.20\linewidth}p{0.05\linewidth}p{0.05\linewidth}p{0.05\linewidth}p{0.2\linewidth}}
    \hline
    \hline
    Cell & Shape       & $d$ (mm) & $l$ (mm) & Coating material & $\gamma_{\mathrm{s}}$ (s$^{-1}$) & $\gamma_{\mathrm{f}}$ (s$^{-1}$) & $N$  & Estimate melting point ($^\circ$C) \\
    \hline
    A    & cylindrical & 51       & 51       & pwMB             & 2.14                             & 24.39                            & 2013 & 60                                 \\
    B    & cylindrical & 51       & 51       & pwMB             & 3.21                             & 27.03                            & 1342 & 60                                 \\
    C    & cylindrical & 51       & 51       & pwMB--H          & 0.74                             & 24.39                            & 5821 & 70                                 \\
    D    & cylindrical & 51       & 51       & pwMB--H          & 1.58                             & 30.30                            & 2726 & 70                                 \\
    E    & spherical   & 25       & ---      & alkene-based     & 3.15                             & 33.33                            & 4184 & 60                                 \\
    F    & spherical   & 30       & ---      & paraffin         & 1.44                             & 23.56                            & 7628 & 60                                 \\
    G    & spherical   & 30       & ---      & paraffin         & 4.35                             & 31.97                            & 2525 & 60                                 \\
    H    & spherical   & 34       & ---      & deuterated       & 43.94                            & ---                              & 221  & $> 100$                            \\
    \hline
    \hline
  \end{tabular}
  \label{tab:cell}
\end{table*}

In order to characterize the high-temperature performance of the cells, careful design of the temperature control system is needed. Since the vapor-phase atoms tend to condense at the coolest part of the cell and therefore form an alkali-metal film on the coating material, it is important to keep the stem temperature lower than the volume temperature. A schematic of the cell thermostat is shown in Fig.~\ref{fig:setup}(b). The vapor cell is housed in an oven made of polyetheretherketone (PEEK) which can be resistively heated with twisted wires. The heating current is controlled with a temperature PID Module (Omega CNi Series) which stabilizes the temperature at the position of TC3. The cell stem is wrapped with a 15~cm long copper braid which is connected to a thermoelectric cooler (TEC) outside of the oven and insulated with low-heat-conductivity foam inside the oven. Cooling the copper braid with the TEC outside the oven creates a temperature gradient and therefore cools the stem of the cell. The temperature can be adjusted by varying the current through the TEC. Temperature measurement is crucial in this experiment. In order to precisely monitor the temperature gradient on the cell, three thermocouples (Tc~1, 2, 3), which are attached to different parts of the cell with high-temperature tape, are used to separately monitor the stem temperature (Tc~1), the capillary temperature (Tc~2) and the volume temperature (Tc~3). An example configuration of the temperature sensors is shown in Fig.~\ref{fig:setup}(c).

\begin{figure}[thbp]
  \centering\includegraphics[]{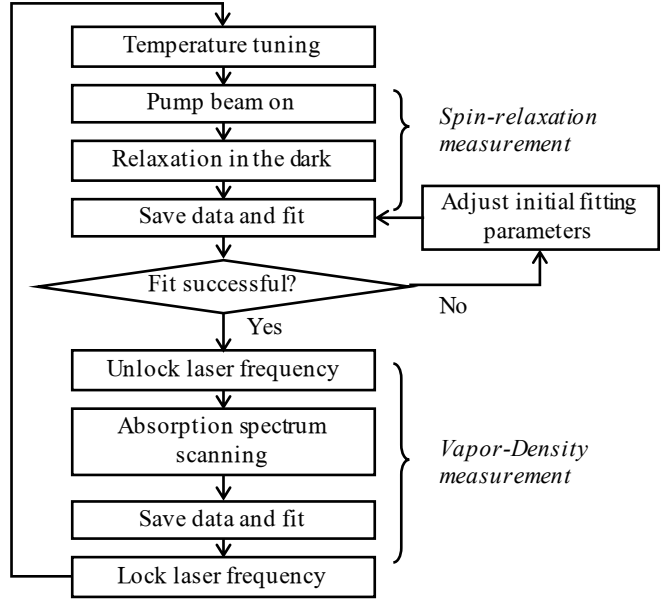}%
  \caption{\label{fig:flow_chart} Simplified flow chart of measurement procedure. At each temperature point, the whole measurement process contains two parts: spin-relaxation measurement and vapor-density measurement. For spin-relaxation measurements, a modified relaxation-in-the-dark method is used and the optical-rotation signal is fitted with a biexponential function. For vapor-density measurements, the spectrum of optical absorption is recorded and fitted with the atomic-absorption profile.}%
\end{figure}

The experimental setup allows for measurements of both the spectrum of optical absorption and the spin-relaxation transient. The measurement procedure is shown in Fig.~\ref{fig:flow_chart}. Once the temperature is tuned, the spin-relaxation transient is measured first. There are two steps in the spin-relaxation measurement. The first step is to pump the atoms in the vapor cell with circularly polarized light. The duration of the pump period is typically 2~s. The second step is to shut the pump light off and let the polarized atoms relax in the dark (except for the weak probe light). For spin relaxation measurements, optical rotation of the probe polarization ($\phi$) is determined using the signals from the polarimeter
\begin{equation}
  \sin{\phi} = \frac{\mathrm{P}_1 - \mathrm{P}_2}{2(\mathrm{P}_1 + \mathrm{P}_2)},
\end{equation}
where $\mathrm{P}_1$ and $\mathrm{P}_2$ correspond to the signals from the two photodiodes in the polarimeter, which are digitized with a 16-bit analog-to-digit converter (ADC) from National Instrument (PCI-6030). Then the derived optical-rotation signal is stored and fitted to a biexponential function\cite{Graf2005}. For vapor-density measurements, the laser frequency is scanned over 20~GHz in 0.5~s by applying a ramp to the DFB current after lifting the frequency lock. A Fabry-P\'erot spectrum analyzer is used to monitor the linearity of the laser-frequency scan. The vapor density can be extracted by fitting the optical-absorption spectrum with the theoretical optical-absorption function. The routines for this are adapted from the AtomicDensityMatrix Mathematica package\cite{ADMpackage}. Before spin-relaxation measurements the laser frequency is locked again.

\section{Results and discussion}

\subsection{Vapor-density measurement}

\begin{figure}[thbp]
  \centering\includegraphics[]{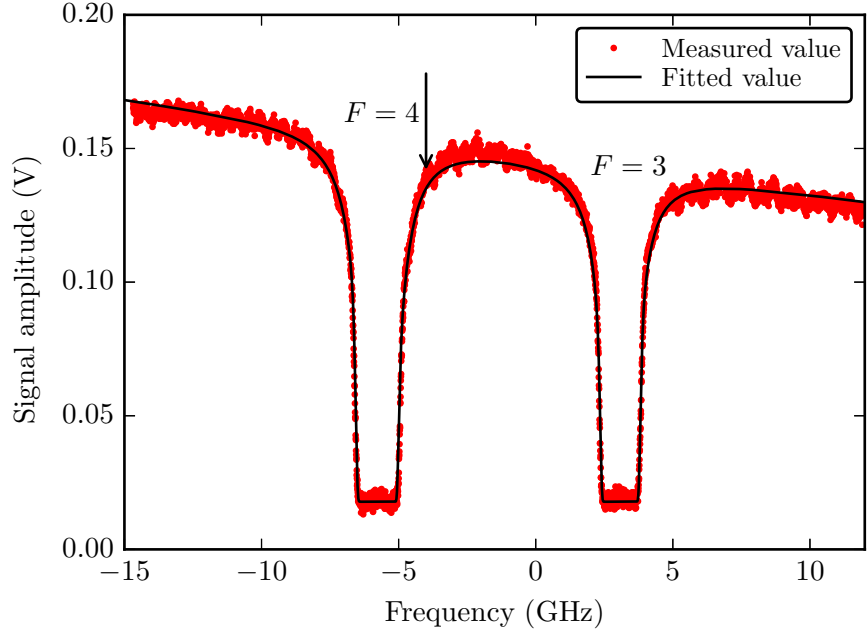}%
  \caption{\label{fig:Absorption_scan} Spectrum of optical absorption
    for Cs D2 line. The data are taken with cell~C at temperatures of
    $T_{\mathrm{stem}}=\SI{82}{\degreeCelsius}$ and
    $T_{\mathrm{volume}}=\SI{90}{\degreeCelsius}$. The fitted vapor
    density is $8.86 \times 10^{12}~\mathrm{cm}^{-3}$. The arrow in
    the figure indicates the locking point of the laser.}%
\end{figure}

Figure~\ref{fig:Absorption_scan} shows an optical-absorption signal of a cell together with the fitting result. The background slope of the absorption signal is due to the laser output power variation during the frequency scan. The nonzero signal value at full absorption is due to an electronic offset. One of the error sources for vapor-density measurement is the nonlinearity of the laser-frequency scan. In the experiment, we use a Fabry-P\'erot spectrum analyzer to linearize the frequency axis. 

\begin{figure}[thbp]
  \centering\includegraphics[]{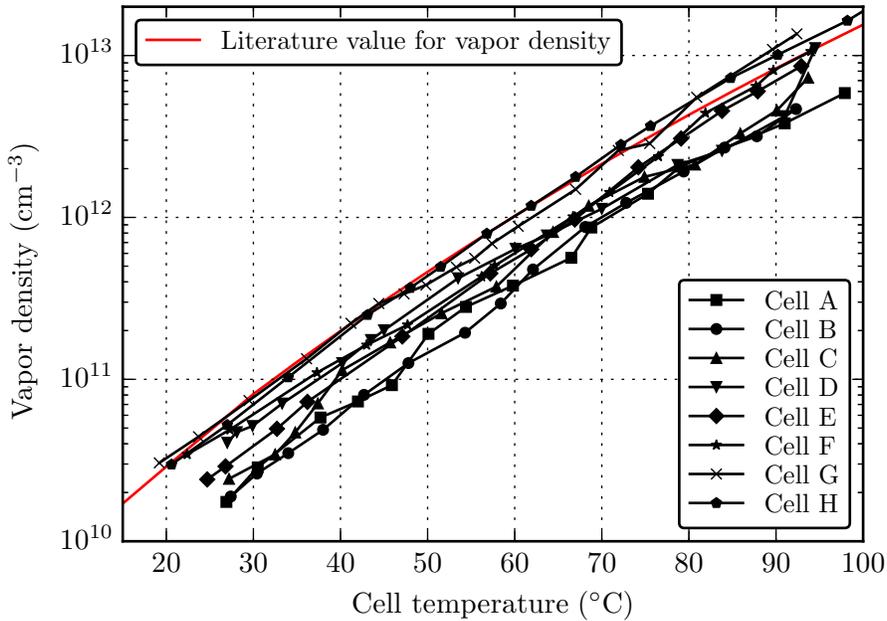}%
  \caption{\label{fig:Vapor_density} Dependence of vapor density on stem temperature. During the measurements, the stem temperature is always kept about $\SI{10}{\degreeCelsius}$ lower than the volume temperature.}%
\end{figure}

The density of the saturated vapor of alkali metal is determined by the lowest temperature in the cell. In the experiment, the stem temperature is kept lower than the volume temperature to guarantee that an alkali-metal film does not form on the inner walls of the working volume of the cell. After the cesium atoms diffuse into the cell volume from the stem and collide with the cell wall, they thermalize with the wall. Therefore, in optical absorption fitting, the Doppler broadening should be determined by the volume temperature. Figure~\ref{fig:Vapor_density} shows the fitted vapor density for several cells as a function of the temperature. The temperature in Fig.~\ref{fig:Vapor_density} is obtained from Tc~2 [see Fig.\ref{fig:setup}(c)]. Due to the finite heat conductivity of glass, there is a temperature gradient along the cell stem. The temperature difference along the stem can be as large as $\SI{10}{\degreeCelsius}$ during the measurement and the temperature gradient can vary from cell to cell. Since the vapor density in AWC cells is normally lower than the saturated vapor density due to the absorption of some alkali atoms by the coating material\cite{Balabas2012,budker2013optical,Seltzer2013book}, we find that the temperature of Tc~2 appears to be the closest to the temperature in the stem that determines the vapor pressure. Moreover, it is also reported that background gases (e.g. $\mathrm{H}_2$, $\mathrm{C}_2\mathrm{H}_4$) can be generated from chemical reactions of alkali atoms and the coating material, which may cause additional pressure broadening in the absorption spectrum\cite{Sekiguchi2016,Seltzer2013book}. However, within the temperature range of our experiment, it was not necessary to include pressure broadening when fitting the spectrum with theoretical optical-absorption function. 

\subsection{Spin-relaxation measurement}
\label{sec:spin-relaxation}

\begin{figure}[thbp]
  \centering\includegraphics[]{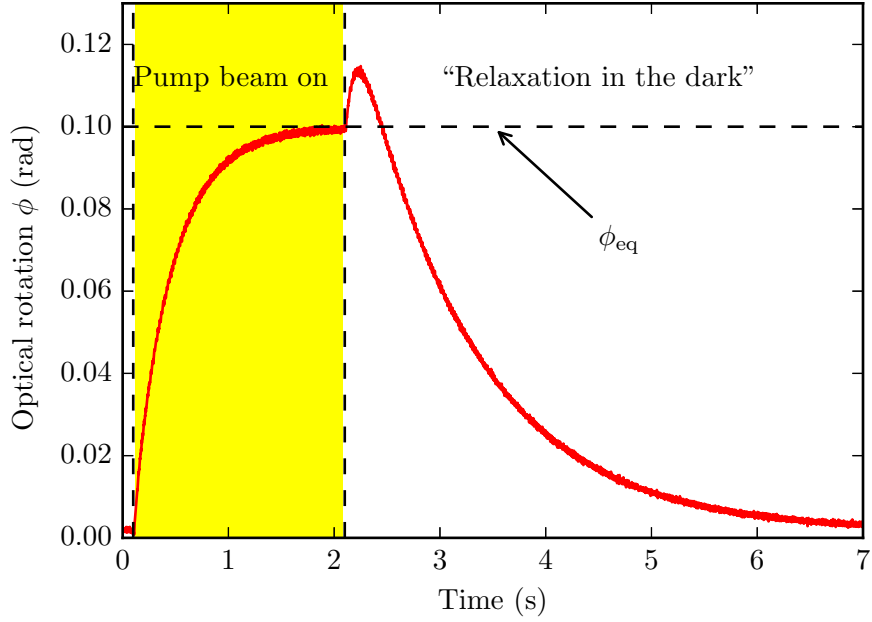}%
  \caption{\label{fig:signal} Optical-rotation signal with Cell~C at room temperature. The signal is obtained with $15~\mathrm{\mu W}$ pump power and $1~\mathrm{\mu W}$ probe power. The curve in the ``relaxation in the dark'' part is fitted with Eq.~(\ref{equ:biexponential}) which gives $\gamma_{\mathrm{s}}=0.927~\mathrm{s}^{-1}$, $\gamma_{\mathrm{f}}=11.1~\mathrm{s}^{-1}$, $A_{\mathrm{s}}=0.137~\mathrm{rad}$, $A_{\mathrm{f}}=-0.042~\mathrm{rad}$, and $\phi_0=0.0018~\mathrm{rad}$.}
\end{figure}

The optical-rotation signal $\phi$ obtained in the spin-relaxation measurement relaxes in a biexponential manner, which can be described with the following formula\cite{Graf2005},
\begin{equation}\label{equ:biexponential}
  \phi = \phi_0 + A_\mathrm{f} \exp{\left(-\gamma_{\mathrm{f}} t\right)} + A_\mathrm{s} \exp{\left(-\gamma_{\mathrm{s}} t\right)},
\end{equation}
where $\phi_0$ is the rotation due to linear effects which is independent of the pump light, $A_{\mathrm{f}}$ and $A_{\mathrm{s}}$ are ``fast'' and ``slow'' relaxation amplitudes and $\gamma_{\mathrm{f}}$ and $\gamma_{\mathrm{s}}$ correspond to ``fast'' and ``slow'' relaxation rates. These two relaxation processes can be related to three primary relaxation mechanisms in the vapor cell: (1)~electron-randomization collisions with the cell wall or gaseous impurities, (2)~spin-exchange collisions between alkali atoms and (3)~exchange of alkali atoms between the volume and the stem, which is also known as the ``uniform relaxation''. The ``fast'' and ``slow'' relaxation rates ($\gamma_{\mathrm{f,s}}$), which are the inverse of the relaxation times ($\tau_{\mathrm{f,s}}$), can be expressed with the following formula\cite{Graf2005},
\begin{equation}
  \label{equ:rel_rates}
  \gamma_{\mathrm{f,s}} = \gamma_{\mathrm{u}} + \frac{1}{64}(33\gamma_{\mathrm{er}} + 22\gamma_{\mathrm{se}}
  \pm\sqrt{961\gamma_{\mathrm{er}}^2+1324\gamma_{\mathrm{er}}\gamma_{\mathrm{se}}+484\gamma_{\mathrm{se}}^2}),
\end{equation}
where $\gamma_{\mathrm{er}}$ is the rate of electron-randomization collisions, $\gamma_{\mathrm{se}}$ is the rate of spin-exchange collisions, and $\gamma_{\mathrm{u}}$ is the rate of uniform relaxation. According to Eq.(\ref{equ:rel_rates}), if $\gamma_{\mathrm{er}} \ll \gamma_{\mathrm{se}}$, the fast and slow relaxation rates are given by
\begin{align}
  \gamma_{\mathrm{f}} & \approx \gamma_{\mathrm{u}} + \gamma_{\mathrm{er}} + \frac{11}{16}\gamma_{\mathrm{se}}, \label{equ:fast_rate}\\
  \gamma_{\mathrm{s}} & \approx \gamma_{\mathrm{u}} + \frac{1}{32}\gamma_{\mathrm{er}}.\label{equ:slow_rate}
\end{align}

Figure~\ref{fig:signal} shows a typical optical rotation signal. When the optical shutter is open during the first 2~s of the measurement, the pump beam creates Zeeman polarization (orientation) in both ground-state hyperfine states and also pumps atoms from the $F=4$ manifold to the $F=3$ manifold producing hyperfine polarization. The relative degree of the two kinds of polarization depends on the frequency tuning of the pump light. Since the probe beam measures the orientation in the $F=4$ manifold, the optical rotation signal increases until it reaches an equilibrium value ($\phi_{\mathrm{eq}}$). In the case that hyperfine polarization is stronger than Zeeman polarization, when the pump beam is turned off and the spin-exchange collisions bring oriented atoms to the $F=4$ manifold, optical rotation experienced by the probe may initially increase before it decreases due to relaxation. Such a situation is seen in Fig.~\ref{fig:signal}.

\begin{figure}[thbp]
  \centering\includegraphics[]{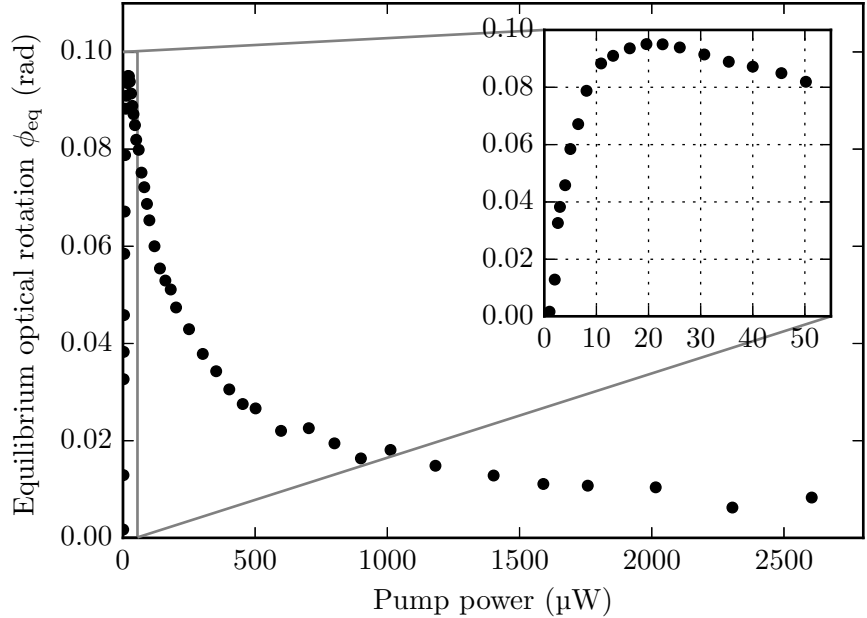}%
  \caption{\label{fig:equilibrium_rotation} Equilibrium optical rotation $\phi_{\mathrm{eq}}$ versus pump power. During the measurement, the probe power is kept at $1~\mathrm{\mu W}$. The data in power-dependence measurements are collected with Cell~C at $T_{\mathrm{stem}}=\SI{18}{\degreeCelsius}$ and $T_{\mathrm{volume}}=\SI{25}{\degreeCelsius}$. The inset shows the equilibrium optical rotation in the low-pump-power regime.}%
\end{figure}

It also should be noted that Eq.(\ref{equ:rel_rates}) is only valid in the condition of low orientation. Since the equilibrium optical rotation ($\phi_{\mathrm{eq}}$), which is shown in Fig.~\ref{fig:signal}, is proportional to the maximum orientation in the $F=4$ manifold created by the pump light, we are able to evaluate the orientation of the atoms in the vapor cell by measuring $\phi_{\mathrm{eq}}$. Figure.~\ref{fig:equilibrium_rotation} shows the dependence of $\phi_{\mathrm{eq}}$ on the pump power. For low pump powers, the orientation in both $F=3$ and $F=4$ manifold increases with pump power so that $\phi_{\mathrm{eq}}$ first increases and reaches its maximum value. The maximum $\phi_{\mathrm{eq}}$ occurs when the pump power is about $25~\mu\mathrm{W}$. Increasing the pump power further decreases $\phi_{\mathrm{eq}}$ because a significant portion of atoms in the $F=4$ manifold is pumped to the $F=3$ manifold. Therefore, we chose $15~\mu\mathrm{W}$ as the pump power to ensure low polarization for all the measurements presented here.

\begin{figure}[thbp]
  \centering\includegraphics[]{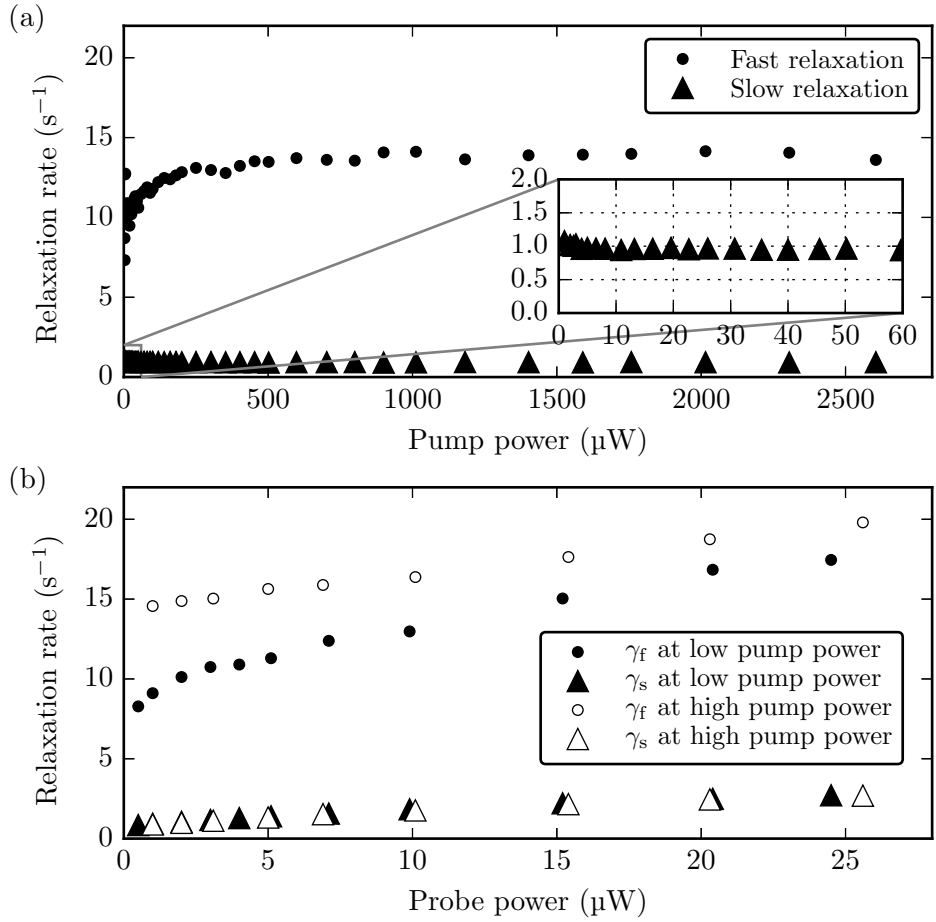}%
  \caption{\label{fig:power_dependence} (a) Dependence of relaxation rates on pump power. During the measurement, the probe power is chosen to be around $1~\mu\mathrm{W}$. The inset shows the slow relaxation rates for low pump powers. For pump power approaching zero, the biexponential fitting gives larger error due to the worse signal-to-noise ratio. The data for the relaxation rates are collected from the same measurement as that for equilibrium optical rotation. (b) Dependence of relaxation rates on probe power. The pump power is $10~\mu\mathrm{W}$ for the low-pump-power case, while it is $2~\mathrm{mW}$ for the high-pump-power case. The probe power is changed by adjusting a neutral density filter.}%
\end{figure}

Figure~\ref{fig:power_dependence}(a) shows the dependence of the relaxation rates on the pump power. The fast relaxation rate stays nearly constant for pump powers above $500~\mu\mathrm{W}$ and reduces for low powers, which can be qualitatively explained by the increase of the slowing-down factor when the atomic polarization is low. The slow relaxation rate appears to be independent of the pump power. The slight increase of the slow relaxation rate at near-zero pump powers is due to the increase of fitting errors. Figure~\ref{fig:power_dependence}(b) shows the dependence of the relaxation rates on the probe power. The dependence is measured with both high pump power ($2~\mathrm{mW}$) and low pump power ($10~\mu\mathrm{W}$). For either case, both fast and slow relaxation rates increase with probe power due to the relaxation of the atomic orientation caused by the linearly polarized probe light. Therefore, in the temperature-dependence measurement, we chose the probe power to be around $1~\mu\mathrm{W}$ to minimize the relaxation due to the probe light while maintaining a sufficient signal-to-noise ratio.

\begin{figure}[thbp]
  \centering\includegraphics[]{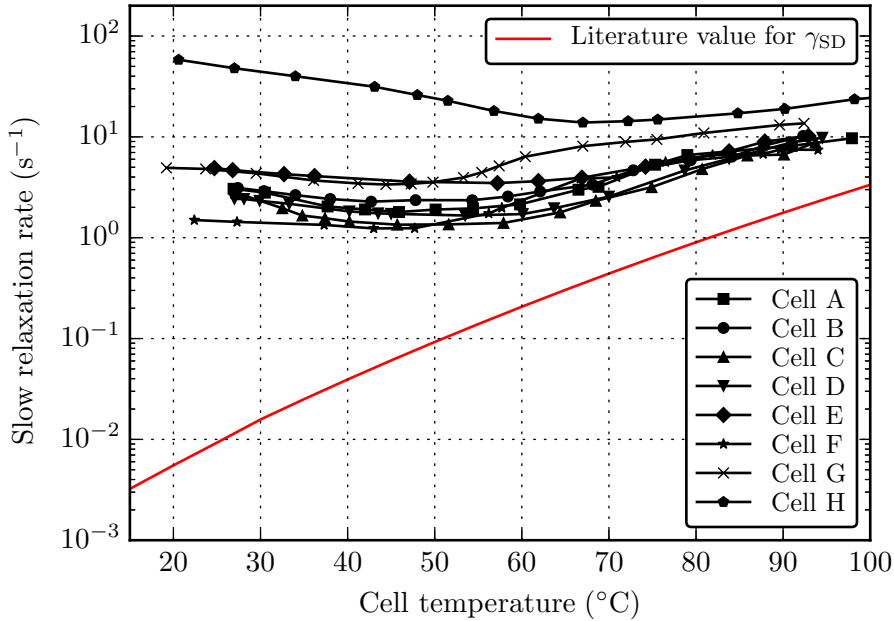}%
  \caption{\label{fig:Relaxation_time} Slow relaxation rates versus temperature. The measured relaxation rates are compared with the literature value for Cs-Cs spin-destruction relaxation rate $\gamma_{\mathrm{SD}}$ [See Section~III~C].}%
\end{figure}

Figure~\ref{fig:Relaxation_time} shows the dependence of the slow relaxation rate on the coating temperature for different cells. In contrast to the vapor density's dependence on the stem temperature the relaxation rate depends mainly on the coating temperature, because the relaxation process is governed by the atom's interaction with the AWC and collisions within the cell volume. For cell~A--G, we observe that as the temperature rises, the slow relaxation rate first decreases then begins to increase at a temperature of around 50-$\SI{60}{\degreeCelsius}$. When the temperature rises above $\SI{70}{\degreeCelsius}$, the slow relaxation rate quickly increases to about 10~s$^{-1}$. Cell~H shows considerably larger slow relaxation rate compared to the other cells at low temperatures; however, it attains comparable performance to the other cells above $\SI{70}{\degreeCelsius}$. The variation of the slow relaxation rate with regard to the temperature change coincides with the drawn-out melting and fusion profiles of pwMB measured with DSC \cite{Seltzer2010}. These show that pwMB is partially molten at temperatures from room temperature to about $\SI{60}{\degreeCelsius}$ and completely melted at temperatures above $\SI{70}{\degreeCelsius}$. This indicates that melting degrades the AWC to some extent. Similar results were reported in~\cite{Bouchiat1966} and~\cite{Balabas2012}. We could also verify that below $\SI{100}{\degreeCelsius}$, where the relaxation rate goes up to around 10~s$^{-1}$, the coating can restore its antirelaxation properties when it is cooled down again. The fast relaxation rate is not shown in the figure because of the large fitting errors due to small signal-to-noise ratios for low beam powers. The amplitude for the fast relaxation becomes indistinguishable from noise at temperatures higher than $\SI{50}{\degreeCelsius}$, so that the transient of the optical rotation signal is well fitted with a single exponential function. According to Fig.~\ref{fig:Relaxation_time}, cells coated with wax distilled at a temperature of $\SI{240}{\degreeCelsius}$ (cell~C and D) show smaller spin relaxation rates than cell~A and B whose coating material was distilled at $\SI{220}{\degreeCelsius}$. As for Cell~E--H, since they have different geometries from those of cylindrical cells, we can compare their coating properties with the others by calculating the average number of bounces without depolarization. The result for the measurement at $T_{\mathrm{volume}} \approx \SI{30}{\degreeCelsius}$ ($T_{\mathrm{stem}} \approx \SI{20}{\degreeCelsius}$) is summarized in
Table~\ref{tab:cell}.

As the cell temperature goes up, the increase of vapor density causes severe degradation of the amplitude of the optical rotation signal. Therefore, in order to guarantee a sufficient signal-to-noise ratio which is crucial for a reliable fit, laser frequency detuning was changed for different temperatures. The laser frequency is locked near high-frequency wing of the absorption profile, which is shown in Fig.~\ref{fig:Absorption_scan}. We have reproduced the results in \cite{Graf2005}, showing that both fast and slow relaxation times are independent of frequency detuning.

\begin{figure}[thbp]
  \centering\includegraphics[]{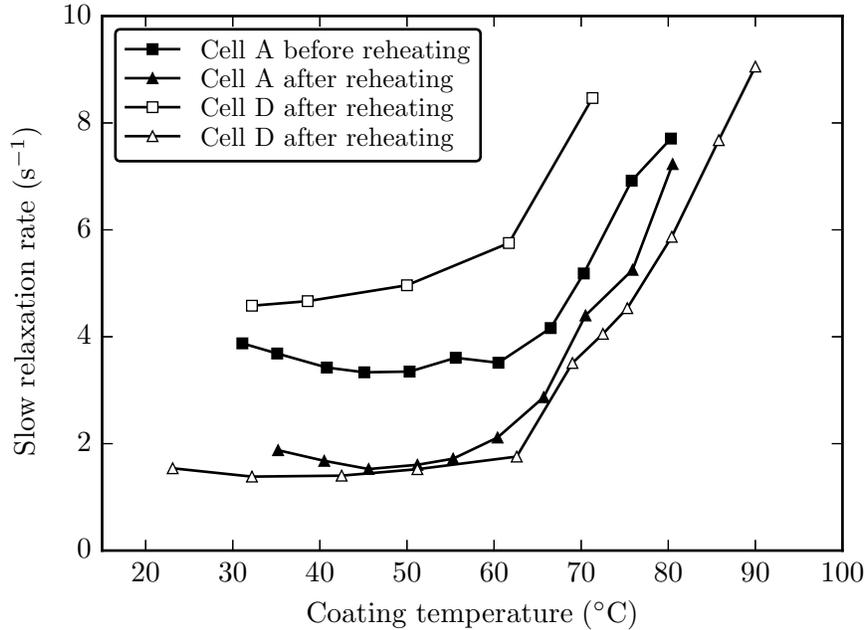}%
  \caption{\label{fig:Reheating} Comparison of slow relaxation rates for cells before and after reheating for cell~A and cell~D. Similar enhancement of AWC relaxation properties by reheating was also observed for other cells.}%
\end{figure}

During the experiment, we observed that spin-relaxation properties can be enhanced for some cells by reheating the cell while ensuring that the stem is, indeed, the coldest part of the cell. Figure~\ref{fig:Reheating} shows the comparison of spin relaxation rates for cell~A and cell~D before and after reheating. For cell~A, the minimum relaxation rate decreased from 3.3~s$^{-1}$ to 1.5~s$^{-1}$. For cell~D, the minimum relaxation rate decreased from 4.5~s$^{-1}$ to 1.3~s$^{-1}$. Similar effects have also been observed for other cells. Reheating involves heating the cell to around $\SI{90}{\degreeCelsius}$ for several hours while keeping the stem temperature a few degrees lower. The reason for this is that, while cells are being stored at room temperature or working at high temperature without maintaining a proper temperature difference between the stem and the volume, alkali atoms tend to form a film or clusters on top of AWC which can cause deterioration of the relaxation properties. By reheating the cell while keeping the stem cooler, the vaporized alkali atoms will only deposit in the stem and the cell appears to be ``cured''.

The validity of the assumption $\gamma_{\mathrm{er}} \ll \gamma_{\mathrm{se}}$ can be checked at both low temperatures and high temperatures. For low temperatures, substituting $\gamma_{\mathrm{f}}$ and $\gamma_{\mathrm{s}}$ in Eq.~(\ref{equ:fast_rate}) and (\ref{equ:slow_rate}) with the measured results from Table.~\ref{tab:cell} and calculating $\gamma_{\mathrm{se}}$ at $T=\SI{30}{\degreeCelsius}$ results in the uniform relaxation rate $\gamma_{\mathrm{u}}$ and the electron-randomization rate $\gamma_{\mathrm{er}}$. Taking cell~D as an example, the spin-exchange relaxation rate can be calculated as $\gamma_{\mathrm{se}} = n_{\mathrm{Cs}} \sigma_{\mathrm{se}} \bar{v} \approx 37.2~\mathrm{s}^{-1}$ where $n_{\mathrm{Cs}} \approx 6 \times 10^{10}~\mathrm{cm}^{-3}$, $\sigma_{\mathrm{se}} = 2\times 10^{-14}~\mathrm{cm}^2$, and $\bar{v} \approx 3.1 \times 10^4~\mathrm{cm}/\mathrm{s}$, so the uniform relaxation rate and the electron-randomization rate can be calculated to be $\gamma_{\mathrm{u}} \approx 1.5~\mathrm{s}^{-1}$ and $\gamma_{\mathrm{er}} \approx 3.2~\mathrm{s}^{-1}$, respectively. For high temperatures, since the fast relaxation rate cannot be reliably measured with the relaxation-in-the-dark method, the upper limit for the electron-randomization rate can be evaluated as $\gamma_{\mathrm{er}} < 32\gamma_{\mathrm{s}}$ [see Eq.~(\ref{equ:slow_rate})]. Again, taking cell~D as an example, the spin-exchange relaxation rate at $\SI{90}{\degreeCelsius}$ can be calculated as $\gamma_{\mathrm{se}} = n_{\mathrm{Cs}} \sigma_{\mathrm{se}} \bar{v} \approx 4760~\mathrm{s}^{-1}$, where $n_{\mathrm{Cs}} \approx 7 \times 10^{12}~\mathrm{cm}^{-3}$ and $\bar{v} \approx 3.4 \times 10^4~\mathrm{cm}/\mathrm{s}$. And the upper limit for the electron-randomization rate is set by $\gamma_{\mathrm{er}} < 32\gamma_{\mathrm{s}}|_{T=\SI{90}{\degreeCelsius}} \approx 213~\mathrm{s}^{-1}$, where $\gamma_{\mathrm{s}}|_{T=\SI{90}{\degreeCelsius}}$ is measured to be about $ 6.67~\mathrm{s}^{-1}$. Based on the above discussion, the assumption that the spin-exchange relaxation rate $\gamma_{\mathrm{se}}$ is much faster than the electron-randomization rate $\gamma_{\mathrm{er}}$ is valid within the whole temperature range of the measurement.

In order to check if there is a significant amount of buffer gas generated after the cells are heated above the melting temperature, we make an order-of-magnitude estimation of how the motion of polarized alkali atoms can be influenced by the background buffer gas. If we assume there is about 5~torr small-molecule buffer gas (H$_2$, CH$_4$, ...) generated at the temperature of $\SI{95}{\degreeCelsius}$, the diffusion coefficient of alkali atoms ($D_{\mathrm{Cs}}$, taking cesium as an example) in the buffer gas is about 400~cm$^2$/s. Then the time for a polarized atom to diffuse over the full length ($l_D$) from the pump beam region to the cell wall (the typical value for $l_D$ can be chosen as 4~cm) is $t = \langle l_D^2 \rangle / (4 D_{\mathrm{Cs}}) \sim 0.01~\mathrm{s}$, which is on the same order of the spin relaxation time we measured at high temperatures. This indicates that if the amount of the generated background buffer gas is larger than 5~torr under such circumstances, by measuring polarized atoms at different locations with respect to the pump beam, we would observe both a significant delay of the response and a variation of the time-dependent signal amplitude in the spin-relaxation measurement. However, for different separations of the pump and the probe beam, neither the delay of the response nor the change of the spin-relaxation amplitude was observed in the experiment. Although the pumping efficiency can be influenced by the optical path length in the cell and the light scattering and refraction at the cell surface, we observed that within the temperature range we explored in the experiment, there is no significant buffer gas generation and the cells still function as the coated cells and not as buffer-gas cells.


\subsection{Figure of merit}

According to Eq.~(\ref{equ:snl}), the spin-projection-noise-limited sensitivity $\delta B_{\mathrm{SNL}}$ for atomic magnetometers is inversely proportional to the square root of the product of spin relaxation time $\tau_{\mathrm{rel}}$ and the number of measured atoms $n$ under a given measurement time. Since polarized atoms can bounce $10^{3} - 10^6$ times from a wall before getting depolarized in AWC cells, the probe beam samples nearly all the atoms in the cell even if the beam size is much smaller than the size of the cell\cite{Zhivun2015}. Therefore, the measured number of atoms in Eq.~(\ref{equ:snl}) can be represented by the total number of atoms $n_{\mathrm{Cs}}$ in the vapor cell. The spin relaxation time $\tau_{\mathrm{rel}}$ is limited by several factors, such as spin-exchange collisions, spin-destruction collisions, wall collisions, optical pumping, and magnetic field gradients. Among them, spin-exchange collisions can be partially or fully suppressed by light-narrowing or SERF techniques. Assuming optical pumping effects and field gradients are sufficiently reduced, the ultimate limit for the relaxation time of vapor-cell magnetometers is set by spin-destruction and wall collisions. Since the slow relaxation rate ($\gamma_{\mathrm{s}}=1/\tau_{\mathrm{s}}$) measured in the experiment is nearly independent of spin-exchange collisions [see Eq.~(\ref{equ:rel_rates})] and dominated by electron-randomization collisions which are essentially spin-destruction and wall collisions, $\tau_{\mathrm{s}}$ can be a good representation for the spin-relaxation time in Eq.~(\ref{equ:snl}) when evaluating the ultimate limit for the sensitivity. As is shown in Fig.~\ref{fig:Relaxation_time}, for increasing temperatures and therefore increasing vapor densities, the slow relaxation rate approaches the spin-destruction limit. Spin-destruction collisions become therefore the dominant mechanism for spin relaxation at high temperatures. The spin-destruction-limited relaxation rate $\gamma_{\mathrm{SD}}$ can be described by the equation $\gamma_{\mathrm{SD}} = (n_{\mathrm{Cs}} \sigma_{\mathrm{SD}} \bar{v})/q$, where $q = 32$ (for Cs) is the scaling factor deduced from Eq.~(\ref{equ:slow_rate}), $n_{\mathrm{Cs}}$ is the measured cesium vapor density, $\sigma_{\mathrm{SD}} = 2\times 10^{-16}~\mathrm{cm}^2$ is the Cs-Cs electron spin-destruction cross section, and $\bar{v} = \sqrt{16 k_{\mathrm{B}} T / (\pi m_{\mathrm{Cs}})}$ is the relative mean thermal velocity for Cs atoms\cite{Bhaskar1980}.

\begin{figure}[thbp]
  \centering\includegraphics[]{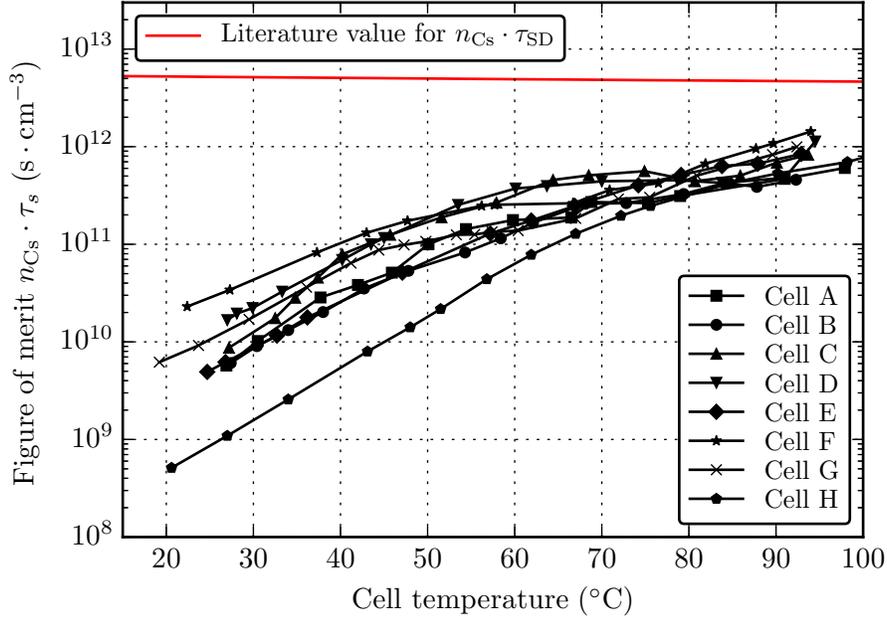}%
  \caption{\label{fig:Figure_of_merit} Dependence of the product of the slow relaxation time $\tau$ and vapor density $n_{\mathrm{Cs}}$ on coating temperature. The stem temperature is kept about $5 - \SI{10}{\degreeCelsius}$ lower than the volume temperature. The literature value is obtained by multiplying Cs-Cs spin-destruction relaxation time with saturated vapor density.}%
\end{figure}

Figure~\ref{fig:Figure_of_merit} shows the product of the spin-relaxation time $\tau_{\mathrm{s}}$ and the vapor density $n_{\mathrm{Cs}}$ for different temperatures. Within the temperature range explored in our experiment, we show that the product of $\tau_{\mathrm{s}}$ and $n_{\mathrm{Cs}}$ increases by a factor of 100 when we increase the temperature from $\SI{25}{\degreeCelsius}$ to $\SI{95}{\degreeCelsius}$. Since the fundamental sensitivity of a Cs magnetometer with a coated cell was shown to be on the order of 1~fT with a 1~s measurement time\cite{Zhivun2015}, considering the benefit from AWC cells operating at high temperatures, we can expect the fundamental sensitivity to be improved to better than 100 aT level. We also notice that at the highest temperature point in our experiment ($T \sim \SI{95}{\degreeCelsius}$), the vapor density reaches $10^{13}~\mathrm{cm}^{-3}$ which already approaches the vapor density typical for SERF magnetometers.

\section{Conclusions}

In conclusion, we investigated the performance of AWC cells by measuring their spin-relaxation rates and vapor densities for different temperatures up to $\SI{95}{\degreeCelsius}$. Previously, AWC (especially paraffin-coated) cells were commonly thought to be unsuitable for high-temperature applications limited by the low melting point of the coating materials. We experimentally observe an increase in the figure of merit as is shown in Fig.~\ref{fig:Figure_of_merit} and demonstrate the potential of AWC cells to work even at temperatures that correspond to partial melting of the coating material, with the potential to improve the sensitivity limit of atomic magnetometers. We note that even though the measured figure of merit is still about one order of magnitude below the spin-destruction limit for the highest temperatures used, the spin-destruction limit will potentially be reached at higher temperatures. Once the figure of merit is ultimately limited by spin-destruction interactions, there will be no more gain in sensitivity with a further increase in vapor density because spin-destruction relaxation time scales as $n^{-1}$. Since the spin-destruction limit is exactly the same limit as is ultimately achieved by SERF magnetometers, we show the possibility of magnetometers based on AWC cells to reach the same sensitivity. And one major advantage of such AWC cells (buffer-gas-free) is that they are much less sensitive to magnetic field gradients\cite{budker2007} and work for a wider range of magnetic fields. Moreover, an additional advantage of magnetometers that work in the high-density regime is their faster response time, which may be helpful for detection of time-varying fields\cite{budker2007,Pustelny2013}. We have focused our discussion on magnetometry applications; however, the results are applicable also to other devices utilizing coated cells, including rotation sensors, electrometers, and secondary frequency standards. Future work includes extending the temperature range of the measurements and experimenting with different coating materials.

\section*{Funding}

National Science Foundation under award CHE-1308381, by the DFG through the DIP program (FO 703/2-1); National Centre for Research and Development within the Leader Program; National Science Fund for Distinguished Young Scholars of China (61225003); National Natural Science Foundation of China (61531003, 61571018); China Scholarship Council (501100004543).

\section*{Acknowledgments}
The author would like to thank E. Zhivun and D. Wurm for continuous help with the experiment and acknowledge A. Shmakov, B. Patton, and V. Heintz for their contributions at the early stages of the project.


\begin{thebibliography}{99}

\bibitem{budker2007} D. Budker and M. Romalis, ``Optical magnetometry,'' Nat. Phys. {\bf 3}, 227 (2007).

\bibitem{Budker2005} D. Budker, L. Hollberg, D. F. Kimball, J. Kitching, S. Pustelny, and V. Yashchuk, ``Microwave transitions and nonlinear magneto-optical rotation in anti-relaxation-coated cells,'' Phys. Rev. A {\bf 71}, 012903 (2005).
\bibitem{Julsgaard2004} B. Julsgaard, J. Sherson, J. I. Cirac, J. Fiurasek, and E. S. Polzik, ``Experimental demonstration of quantum memory for light,'' \nat {\bf 71}, 482 (2004).

\bibitem{Kuzmich2000} A. Kuzmich, L. Mandel, and N. P. Bigelow, ``Generation of spin squeezing via continuous quantum nondemolition measurement,'' \prl {\bf 85}, 1594 (2000).

\bibitem{Pustelny2013} S. Pustelny, D. F. Jackson Kimball, C. Pankow, M. P. Ledbetter, P. Wlodarczyk, P. Wcislo, M. Pospelov, J. R. Smith, J. Read, W. Gawlik, and D. Budker, ``The Global Network of Optical Magnetometers for Exotic physics (GNOME): A novel scheme to search for physics beyond the Standard Model,'' Ann. Phys. {\bf 525}, 659 (2013).

\bibitem{Pustelny2006} S. Pustelny, D. Jackson Kimball, S. Rochester, V. Yashchuk, and D. Budker, ``Influence of magnetic-field inhomogeneity on nonlinear magneto-optical resonances,'' \pra {\bf 74}, 063406 (2006).

\bibitem{Bouchiat1966} M. -A. Bouchiat and J. Brossel, ``Relaxation of optically pumped Rb atoms on paraffin-coated palls,'' Phys. Rev. {\bf 147}, 41 (1966).

\bibitem{Liberman1986} V. Liberman and R. J. Knize, ``Relaxation of optically pumped Cs in wall-coated cells,'' \pra {\bf 34}, 5115 (1986).

\bibitem{Graf2005} M. Graf, D. Kimball, S. Rochester, K. Kerner, C. Wong, D. Budker, E. Alexandrov, M. Balabas, and V. Yashchuk, ``Relaxation of atomic polarization in paraffin-coated cesium vapor cells,'' \pra {\bf 72}, 023401 (2005).

\bibitem{Karaulanov2009} T. Karaulanov, M. Graf, D. English, S. Rochester, Y. Rosen, K. Tsigutkin, D. Budker, E. Alexandrov, M. V. Balabas, D. F. Kimball, F. Narducci, S. Pustelny, and V. Yashchuk, ``Controlling atomic vapor density in paraffin-coated cells using light-induced atomic desorption,'' \pra {\bf 79}, 012902 (2009).

\bibitem{Balabas2010} M. V. Balabas, K. Jensen, W. Wasilewski, H. Krauter, L.S. Madsen, J.H. M\"uller, T. Fernholz, and E.S. Polzik, ``High quality anti-relaxation coating material for alkali atom vapor cells,'' \opex {\bf 18}, 5825 (2010).

\bibitem{Balabas2010a} M. V. Balabas, ``Dependence of the longitudinal relaxation time of the polarization of cesium atoms in the ground state on the temperature of an antirelaxation coating,'' Tech. Phys. {\bf 55}, 1324 (2010).

\bibitem{Balabas2012} M. V. Balabas and O. Y. Tret'yak, ``Temperature dependence of the kinetics of irreversible escape of cesium atoms from a vapor phase into an antirelaxation coating,'' Tech. Phys. {\bf 57}, 1257 (2012).

\bibitem{Nasyrov2015} K. Nasyrov, S. Gozzini, A. Lucchesini, C. Marinelli, S. Gateva, S. Cartaleva, and L. Marmugi, ``Antirelaxation coatings in coherent spectroscopy: Theoretical investigation and experimental test,'' \pra {\bf 92}, 043803 (2015).

\bibitem{Seltzer2010} S. J. Seltzer, D. J. Michalak, M. H. Donaldson, M. V. Balabas, S.K. Barber, S. L. Bernasek, M. -A. Bouchiat, A. Hexemer, A. M. Hibberd, D. F. Kimball, C. Jaye, T. Karaulanov, F. A. Narducci, S. A. Rangwala, H. G. Robinson, A. K. Shmakov, D.L. Voronov, V. V. Yashchuk, A. Pines, and D. Budker, ``Investigation of antirelaxation coatings for alkali-metal vapor cells using surface science techniques,'' J. Chem. Phys. {\bf 133}, 144703 (2010).

\bibitem{Balabas2010PRL} M. V. Balabas, T. Karaulanov, M. P. Ledbetter, and D. Budker, ``Polarized alkali-metal vapor with minute-long transverse spin-relaxation time,'' \prl{\bf 105}, 070801 (2010).

\bibitem{Balabas2013} M. V. Balabas and O. Y. Tretiak, ``Comparative study of alkali-vapour cells with alkane-, alkene- and 1-nonadecylbenzene-based antirelaxation wall coatings,'' Quantum Electron. {\bf 43}, 1175 (2013).

\bibitem{Jensen2016} K. Jensen, R. Budvytyte, R. A. Thomas, T. Wang, A. Fuchs, M. V. Balabas, G. Vasilakis, L. Mosgaard, T. Heimburg, S. Olesen, and E.S. Polzik, ``Non-invasive detection of animal nerve impulses with an atomic magnetometer operating near quantum limited sensitivity,'' arXiv:1601.03273 (2016).

\bibitem{Kitching2002} J. Kitching, S. Knappe, and L. Hollberg, ``Miniature vapor-cell atomic-frequency references,'' \apl {\bf 81}, 553 (2002).

\bibitem{Ledbetter2008} M. P. Ledbetter, I. M. Savukov, V. M. Acosta, D. Budker, and M. V. Romalis, ``Spin-exchange-relaxation-free magnetometry with Cs vapor,'' \pra {\bf 77}, 1 (2008).

\bibitem{Seltzer2009} S. J. Seltzer and M. V. Romalis, ``High-temperature alkali vapor cells with antirelaxation surface coatings,'' J. Appl. Phys. {\bf 106}, 114905 (2009).

\bibitem{Rahman1987} C. Rahman and H. G. Robinson, ``Rb O-O hyperfine transition in evacuated wall-coated cell at melting temperature,'' IEEE J. Quantum Electron. {\bf 23}, 452 (1987).

\bibitem{Vanier1981} J. Vanier, R. Kunski, A. Brisson, and P. Paulin, ``Progress and prospects in rubidium frequency standards,'' J. Phys. {\bf 42}, 139 (1981).

\bibitem{budker2013optical} D. Budker and D. F. Kimball, {\it Optical Magnetometry} (Cambridge University, 2013).

\bibitem{footnote1} It is often the transverse spin relaxation time ($T_2$) that appears in Eq.~(\ref{equ:snl}) to represent the spin relaxation time $\tau_{\mathrm{rel}}$. Since the transverse spin relaxation time is limited by the longitudinal spin relaxation time ($T_1$) ($T_2 \le 2T_1$, which is the regime where atomic magnetometers are often operated), we use $T_1$ to represent the spin-relaxation time $\tau_{\mathrm{rel}}$ in Eq.~(\ref{equ:snl}) to study the fundamental limit to the magnetometric sensitivity.

\bibitem{Franzen1959} W. Franzen, ``Spin relaxation of optically aligned rubidium vapor,'' Phys. Rev. {\bf 115}, 850 (1959).

\bibitem{Chalupczak2013} W. Chalupczak, P. Josephs-Franks, R. M. Godun, and S. Pustelny, ``Radio-frequency spectroscopy in the dark,'' \pra {\bf 88}, 052508 (2013).

\bibitem{Yashchuk2000} V. V. Yashchuk, D. Budker, and J. R. Davis, ``Laser frequency stabilization using linear magneto-optics,'' Rev. Sci. Instrum. {\bf 71}, 341 (2000).

\bibitem{footnote2} Even larger locking ranges may be achieved with DAVLLs operated at alleviated temperatures\cite{Marchant2011} or systems using buffer-gas cells\cite{Pustelny2015}.

\bibitem{ADMpackage} S. Rochester, {\it AtomicDensityMatrix} v15.08.27 (2015).

\bibitem{Seltzer2013book} S. J. Seltzer, M. -A. Bouchiat, and M. V. Balabas, ``Surface coatings for atomic magnetometry,'' in {\it Optical Magnetometry}, D. Budker and D. F. Kimball, ed. (Cambridge University, 2013) Chap. 11, pp. 205--224.

\bibitem{Sekiguchi2016} N. Sekiguchi and A. Hatakeyama, ``Non-negligible collisions of alkali atoms with background gas in buffer-gas-free cells coated with paraffin,'' Appl. Phys. B {\bf 122}, 81 (2016).

\bibitem{Zhivun2015} E. Zhivun, A. Wickenbrock, J. Sudyka, S. Pustelny, B. Patton, and D. Budker, ``Light shift averaging in paraffin-coated alkali vapor cells,'' arXiv:1511.05345 (2015).

\bibitem{Bhaskar1980} N. Bhaskar, J. Pietras, J. Camparo, W. Happer, and J. Liran, ``Spin destruction in collisions between cesium atoms,'' Phys. Rev. Lett. {\bf 44}, 930 (1980).

\bibitem{Marchant2011} A. L. Marchant, S. H\"andel, T. P. Wiles, S. A. Hopkins, C. S. Adams, and S. L. Cornish, ``Off-resonance laser frequency stabilization using the Faraday effect,'' Opt. Lett. {\bf 36}, 64 (2011).

\bibitem{Pustelny2015} S. Pustelny, V. Schultze, T. Sholtes, and D. Budker, ``Dichroic atomic vapor laser lock with multi-gigahertz stabilization range,'' arXiv:1512.08919 (2015).

\bibitem{Zhivun2014} E. Zhivun, A. Wickenbrock, B. Patton, and D. Budker, ``Alkali-vapor magnetic resonance driven by fictitious radiofrequency fields,'' \apl {\bf 105}, 192406 (2014).
  
\end{thebibliography}
\end{document}